\documentclass{llncs}
\usepackage{import}

\usepackage{ifthen}
\usepackage{url}
\usepackage{amssymb}
\usepackage{amsmath}
\usepackage{import}
\usepackage{xparse}
\usepackage{amsmath}
\usepackage[usenames,dvipsnames]{xcolor}
\usepackage{xspace}
\usepackage{datatool}
\usepackage{glossaries}

\newcommand{\m}[1]{\ensuremath{#1}\xspace}

	\newcommand{\limplies}{\m{\Rightarrow}}
	\newcommand{\limpl}{\limplies}
	\newcommand{\lequiv}{\m{\Leftrightarrow}}

	\newcommand{\lrule}{\m{\leftarrow}}
	\newcommand{\cause}{\m{\stackrel{c}{\lrule}}}

	\newcommand{\voc}{\m{\Sigma}}

	\newcommand{\struct}{\m{I}}

	\newcommand{\theory}{\m{\mathcal{T}}}

	\NewDocumentCommand\inter{g+g}{%
	  \IfNoValueTF{#1}
	    {\struct}
	    {\m{#1^{#2}}}}
	
	\newcommand{\model}{\m{M}}

	\newcommand{\bool}{\m{\mathbb{B}}}

	\renewcommand{\int}{\m{\mathbb{Z}}}

	\NewDocumentCommand\subs{g+g}{%
	  \IfNoValueTF{#1}
	    {\m{/}}
	    {\m{#1/ #2}}}

	\newcommand{\logicname}[1]{\text{\sc #1}\xspace}

	\newcommand{\idp}{\logicname{IDP}}

	\newcommand{\idpdthree}{\logicname{IDPD3}}
	\newcommand{\idpwebide}{\logicname{IDP Web-IDE}}
	\newcommand{\idpdraw}{\logicname{ID$^{P}_{Draw}$}}

	\newcommand{\fodotidp}{\logicname{FO(\ensuremath{\cdot})\ensuremath{^{\mathtt{IDP}}}}}
	\newcommand{\foidp}{\fodotidp}

\newcommand{\ouracronym}[3]{%
	\newacronym{#1}{#2}{#3}
	\expandafter\newcommand\csname #1\endcsname{\gls{#1}\xspace}%
}
	\ouracronym{FO}{FO}{first-order logic}
	\ouracronym{MX}{MX}{Model Expansion}
	\ouracronym{MO}{MO}{Model Optimization}
	\ouracronym{ASP}{ASP}{Answer Set Programming}
	\ouracronym{TP}{TP}{Theorem Proving}
	\ouracronym{LP}{LP}{Logic Programming}
	\ouracronym{CP}{CP}{Constraint Programming}
	\ouracronym{FP}{FP}{Functional Programming}
	\ouracronym{KR}{KR}{Knowledge Representation}
	\ouracronym{CSP}{CSP}{Constraint Satisfaction Problem}
	\ouracronym{SMT}{SMT}{SAT Modulo Theories}
	\ouracronym{KBS}{KBS}{Knowledge Base System}
	\ouracronym{NNF}{NNF}{Negation Normal Form}
	\ouracronym{FNNF}{FNNF}{Flat Negation Normal Form}
	\ouracronym{DefNNF}{DefNNF}{Definition Negation Normal Form}
	\ouracronym{CDCL}{CDCL}{Conflict-Driven Clause-Learning}
	\ouracronym{LCG}{LCG}{Lazy Clause Generation}

	\makeatletter
	\def\ifenv#1{
	\def\@tempa{#1}%
	\def\@ttempa{#1*}%
	\ifx\@tempa\@currenvir
	\expandafter\@firstoftwo
	\else
	\expandafter\@secondoftwo
	\fi
	}
	\makeatother

	\newcommand{\ddrule}[4]{\ensuremath{#1 \leftarrow #2 & \{#3\} & #4}}
	\newcommand{\drule}[2]{\ensuremath{#1 & \leftarrow & #2}}

	\newcommand{\darule}[4]{\ensuremath{#1 \leftarrow #2 & \{#3\} & #4}}
	\newcommand{\arule}[2]{\ensuremath{#1 \, &\leftarrow \, #2}}

	\newcommand{\LNDRule}[2]{
	\ifenv{array}
	{\drule{#1}{#2}}
	{ \ifenv{align}
		{\arule{#1}{#2}}
		{\ifenv{align*}
		{\arule{#1}{#2}}
		{ERROR: using LDRule in unsupported environment: \@currenvir}
		}
	}
	}

	\newcommand{\LDRule}[4]{
	\ifenv{array}
	{\ddrule{#1}{#2}{#3}{#4}}
	{ \ifenv{align}
		{\darule{#1}{#2}{#3}{#4}}
		{\ifenv{align*}
		{\darule{#1}{#2}{#3}{#4}}
		{ERROR: using LDRule in unsupported environment: \@currenvir}
		}
	}
	}

	\NewDocumentCommand\LRule{m+g+g+g}{%
		\IfNoValueTF{#2}%
		{#1.&}{%
		\IfNoValueTF{#3}
		{\LNDRule{#1}{#2.}}
		{\LDRule{#1}{#2.}{#3}{#4}}%
		}
	}

	\NewDocumentCommand\CLRule{m+g}{%
	\ifenv{array}
	{\cdrule{#1}{#2}}
	{ \ifenv{align}
		{\carule{#1}{#2}}
		{\ifenv{align*}
			{\carule{#1}{#2}}
			{ERROR: using CLRule in unsupported environment: \@currenvir}
		}
	}
	}

	\NewDocumentCommand\carule{m+g}{%
		\IfNoValueTF{#2}
			{\ensuremath{#1.}}
			{\ensuremath{#1 \, &\cause \, #2}}}
	\NewDocumentCommand\cdrule{m+g}{%
		\IfNoValueTF{#2}
			{\ensuremath{#1.}}
			{\ensuremath{#1 & \cause & #2}}}

	\newcommand{\algrule}[4]{
	\hbox{{#1}:}& 
	\quad #2 ~\longrightarrow~ #3 
	\hbox{~ if } #4\\
	}

	\newcommand{\AlgoRule}[4]{
	\ifenv{array}
	{\algrule{#1}{#2}{#3}{#4}}
		{ERROR: using AlgoRule in unsupported environment: \@currenvir}
	}

\newcommand{\commentstyle}{\color{Gray}}

	\usepackage[final]{listings}

	\lstdefinelanguage{idp}{
		morekeywords=[1]{namespace,vocabulary,theory,structure,procedure,term,set,formula, spec, specification},
		morekeywords=[2]{include,using,type,isa,contains,partial,extern,LFD,GFD,constructed,from,constraint,func,pred,supertype,of,subtype,define},
		morekeywords=[3]{int,float,char,string,nat},
		morekeywords=[4]{if,then,else,for,end},
		morecomment=[s]{/*}{*/},	
		morecomment=[l]{//}
	}
	\newcommand{\ignore}[1]{}

	\newboolean{nocomments}
	\setboolean{nocomments}{false}

	\newboolean{commentmargin}
	\setboolean{commentmargin}{true}

	\newcommand{\namedcomment}[3]{
		\ifthenelse{\boolean{nocomments}}
		{} 
		{ 
			\ifthenelse{\boolean{commentmargin}}
				{ {\color{#3} \marginpar{\color{#3}\sc #2}#1}  } 
				{  {\color{#3} {\sc #2}: #1}  } 
		}
	}
	\newcommand{\mnamedcomment}[3]{\ifthenelse{\boolean{nocomments}}{}{{\marginpar{ \color{#3}{\sc #2}:#1}}}}

	\usepackage{soul}

\usepackage{etoolbox}

\newcommand\setcitation[2]{%
  \csdef{mycommoncitation#1}{#2}}

\setcitation{IDP}{WarrenBook/DeCatBBD14}
\setcitation{fodot}{tocl/DeneckerT08}
\setcitation{CP}{Apt03}
\setcitation{EZCSP}{lpnmr/Balduccini11}
\setcitation{KR}{Baral:2003}
\setcitation{ASPComp2}{lpnmr/DeneckerVBGT09}
\setcitation{ASPComp3}{journals/tplp/CalimeriIR14}
\setcitation{ASPComp4}{conf/lpnmr/AlvianoCCDDIKKOPPRRSSSWX13}
\setcitation{CPSupport}{ictai/DeCat13}
\setcitation{functionDetection}{iclp/DeCatB13}
\setcitation{fodot2asp}{DeneckerLTV12} 
\setcitation{Tarskian}{DeneckerLTV12} 
\setcitation{Inca}{iclp/DrescherW12}
\setcitation{csp2asp}{ijcai/DrescherW11}
\setcitation{DPLLT}{cav/GanzingerHNOT04}
\setcitation{AspInPractice}{synthesis/2012Gebser}
\setcitation{clasp}{ai/GebserKS12}
\setcitation{oclingo}{kr/GebserGKOSS12}
\setcitation{gringo}{lpnmr/GebserST07}
\setcitation{cmodels}{aaai/GiunchigliaLM04}
\setcitation{inputster}{tplp/Jansen13}
\setcitation{DLV}{tocl/LeonePFEGPS06}
\setcitation{LearningPaper}{TPLP/BruynoogheBBDDJLRDV} 
\setcitation{clog}{iclp/BogaertsVDV14} 
\setcitation{foc}{iclp/BogaertsVDV14}
\setcitation{inferenceClog}{ecai/BogaertsVDV14}
\setcitation{examplesClog}{nmr/BogaertsVDV14b} 
\setcitation{AFT}{DeneckerBV12}
\setcitation{KBS}{iclp/DeneckerV08}
\setcitation{KBPE}{inap/DePooterWD11}
\setcitation{lazyGrounding}{corr/CatDSB14} 
\setcitation{ASP}{marek99stable}
\setcitation{satid}{sat/MarienWDB08}
\setcitation{lazyclausegeneration}{constraints/OhrimenkoSC09}
\setcitation{FP}{ACMCS/Hudak89}
\setcitation{GroundingWithBounds}{jair/WittocxMD10}
\setcitation{SAT}{faia/SilvaLM09}
\setcitation{LTC}{iclp/Bogaerts14}
\setcitation{SPSAT}{ictai/DevriendtBMDD12}
\setcitation{BreakID}{cspsat/DevriendtBB14}
\setcitation{LCG}{stuckeyLCG}
\setcitation{MiniZinc}{conf/cp/NethercoteSBBDT07}
\setcitation{amadini}{cpaior/AmadiniGM13}
\setcitation{bootstrapping}{lash/BogaertsJDJBD14}
\setcitation{GroundedFixpoints}{ai/BogaertsVD15}
\setcitation{LogicBlox}{datalog/GreenAK12}
\setcitation{proB}{journals/sttt/LeuschelB08}
\setcitation{NaturalInductions}{KR/DeneckerV14} 
\setcitation{LP}{jacm/EmdenK76}
\setcitation{SMT}{faia/BarrettSST09}
\setcitation{AF}{ai/Dung95}
\setcitation{ADF}{kr/BrewkaW10}
\setcitation{ADFRevisited}{ijcai/BrewkaSEWW13}
\setcitation{DefaultLogic}{ai/Reiter80}
\setcitation{DL}{ai/Reiter80}
\setcitation{AEL}{mo85}
\setcitation{}{}
\setcitation{completion}{adbt/Clark78}
\setcitation{}{}
\setcitation{}{}
\setcitation{}{}
\setcitation{}{}
\setcitation{}{}
\setcitation{}{}
\setcitation{}{}
\setcitation{}{}
\setcitation{}{}
\setcitation{}{}
\setcitation{}{}
\setcitation{}{}
\setcitation{}{}
\setcitation{}{}
\setcitation{}{}
\setcitation{}{}
\setcitation{}{}
\setcitation{}{}
\setcitation{}{}
\setcitation{}{}
\setcitation{}{}
\setcitation{}{}
\setcitation{}{}
\setcitation{}{}
\setcitation{}{}
\setcitation{}{}
\setcitation{}{}
\setcitation{}{}
\setcitation{}{}
\setcitation{}{}
\setcitation{}{}
\setcitation{}{}
\setcitation{}{}
\setcitation{}{}
\setcitation{}{}
\setcitation{}{}
\setcitation{}{}
\setcitation{}{}
\setcitation{}{}
\setcitation{}{}
\setcitation{}{}
\setcitation{}{}
\setcitation{}{}
\setcitation{}{}
\setcitation{}{}
\setcitation{}{}
\setcitation{}{}
\setcitation{}{}
  
\title{Visualising interactive inferences with \IDPD}
\author{Ruben Lapauw, Ingmar Dasseville, Marc Denecker}
\institute{KU Leuven} 
\date{\today}
\usepackage[
    style=numeric,
    sorting=nty,
    doi=false, 
    backend=bibtex,
    sortcites=true,
    maxnames=10
    ]{biblatex}
\usepackage[utf8]{inputenc}
\usepackage{graphicx}
\usepackage[]{hyperref}
\usepackage{capt-of}
\newcommand{\IDP}{\idp}

\newcommand{\FOIDP}{\foidp}
\newcommand{\IDPDRAW}{\idpdraw}
\newcommand{\entails}{\vDash}

\newcommand{\IDPD}{\idpdthree}
\newcommand{\IDPW}{\idpwebide}
\newcommand{\ASPVIZ}{\m{ASPVIZ}}
\newcommand{\citeFull}[1]{\citeauthor{#1}~\cite{#1}}
\ouracronym{DSL}{DSL}{domain specific language}
\usepackage{newfloat}
\usepackage{caption}
\DeclareFloatingEnvironment[fileext=frm,
	placement={!ht},name=Listing,within=none]{flisting}
\newcommand{\linkLesrooster}{\url{https://dtai.cs.kuleuven.be/krr/idp-ide/?present=Roster}}
\newcommand{\linkPacMan}{\url{http://dtai.cs.kuleuven.be/krr/idp-ide/?present=pacman}}
\newcommand{\linkMinimalExample}{\url{https://dtai.cs.kuleuven.be/krr/idp-ide/?present=Count}}

\begin{document}
\maketitle

\begin{abstract}
A large part of the use of knowledge base systems is the interpretation of the output by the end-users and the interaction with these users. 
Even during the development process visualisations can be a great help to the developer.
We created \IDPD as a library to visualise models of logic theories. \IDPD is a new version of \IDPDRAW  and adds support for visualised interactive simulations.
\end{abstract}

\section{Introduction} 
Current logic inference systems often communicate with the user in a text-based method.
Interpreting output of the system, e.g. in the form of a structure of an answer set or a structure, is often difficult for the user.
Appropriate visualisations of the output can be an enormous help to the user.
However creating a fitting visualisation is a cumbersome task.
Therefore in the past, several system were developed to build visualisations  with logic inference engines where logic itself is used to describe the visualisation (e.g. Kara~\cite{kloimullnerkara},\ASPVIZ~\cite{cliffe2008aspviz} and \IDPDRAW~\cite{url:idpdraw}).

These systems generally take an answer set containing special graphical facts. 
The combination of these facts is interpreted by the system to produce the visualisation. 
Kara extends this approach with a generalized visualisation approach that can visualise any answer set as a hypergraph. 
Kara also allows some interaction with the user by using abductive reasoning. 
This abductive reasoning will modify the initial interpretation to match the visualised output. 
This interactivity however cannot be used as a simulation. 

\IDPD is the successor of \IDPDRAW. Unlike \IDPDRAW which was written in C++, \IDPD is written in Lua and Javascript with supporting libraries in both languages: a JSON encoder for Lua and the d3 visualisation library~\cite{2011-d3}.

\IDPD is now fully integrated with \IDP and the \IDPW.
This change allows for a better portability and maintenance.
It also allows easier development of applications.

\IDPD keeps most of the original features like drawing multiple frames of rectangles and circles. 
It also keeps the possibility to visualise graphs with an automatically generated layout.
However some functions like polygons are not yet implemented.
The original system is extended with basic animations when transitioning between two frames.
And more importantly \IDPD has support for interactively visualised linear time calculus theories.

In \citeFull{iclp/Bogaerts14} a system is described that takes as input a logic linear time calculus theory that describes a domain of interactive processes. The system then simulates an interactive process satisfying this theory using in a loop of user input and the progression inference to decide the next state. 
So far this system was text-based. 
The \IDPD library allows to build a graphical interface to this system, visualising the current state and capturing user interaction with this visualisation to determine the next state.

In the following sections we describe \IDP as a knowledge base system, introduce \IDPD with its features and implementation and show how to create a full interactive simulation of a linear time theory.
After this we will compare \IDPD to other visualisation systems and describe some future extensions.

\section{\IDP}
Input to \IDP typically consists of four types of objects: a vocabulary (\voc), a theory (\theory), a structure ($S$) and procedures. The first three declare the knowledge. The last is an imperative method to run inferences on the declared knowledge. An example of each can be found in Listing~\ref{lst:input}.

A vocabulary object has the syntactical form: \[\texttt{vocabulary <name> \{ <list of symbols>\}}\] This declares a named vocabulary consisting of a set of symbols: types, typed predicates and typed functions. 

A theory object has the syntactical form: \[\texttt{theory <name> : <voc name> \{ <list of sentences> \}}\] This form declares a named theory over a vocabulary with a given set of sentences. The symbols used in the sentences must be a part of the vocabulary.
The language of sentences in \IDP supports the classical logical operators ($\land$, $\lor$, $\lnot$, $\limplies$, $\limpl$, $\lequiv$). \IDP extends this with predicates and quantifications (\mbox{$\forall x/\exists x : \phi(x)$}), with functions, arithmetic and aggregates ($sum\{x : P(x) : x\}$) for numeric support, with (inductive) definitions (sets of rules, calculated under the well-founded semantics~\cite{pods/GelderRS88}) and type derivation and checking. This language of \IDP is an extension of \FO named \FOIDP.

A structure object has the syntactical form \[\texttt{structure <name> : <voc name> \{ <list of interpretations> \}}\]
The structure must contain a full interpretation of the types of the vocabulary (the domain) and a partial interpretation of the other symbols in the vocabulary. A structure is called two-valued if every symbol is fully interpreted. A model is a structure that satisfies a theory.

A procedure object has the syntactical form \[\texttt{procedure <name>(<parameters>) \{<instructions>\}}\]
A procedure has a name, parameters and imperative instructions in Lua. Lua is used as a scripting language in \IDP to execute inferences on the above logic objects. Inferences are special procedures built in \IDP to reason with these vocabularies, theories and structures (cf.  \citeFull{WarrenBook/DeCatBBD14}). In this paper only two of these inferences are used: model expansion and progression~\cite{iclp/Bogaerts14}. 

\subsubsection{Modelexpansion($T$, $S$)}
The first inference, model expansion, expands a partial structure \struct to a model \model so that $\struct \subseteq \model$ and $\model \entails \theory$. 

For example the vocabulary(V), theory(T) and structure(S) from Listing~\ref{lst:input} can be expanded to the three models (M1, M2, M3) from Listing~\ref{lst:output}:
\begin{center}
\begin{minipage}{0.4\textwidth}
\begin{lstlisting}
vocabulary V : {
  type Num isa int
  A : Num
  B : Num
}
theory T : V {
  A + B > 8.
}
structure S : V {
  Num = {1..5}
}
procedure main() {
  stdoptions.nbmodels=4;
  printmodels(modelexpand(T, S))
}
\end{lstlisting}
\captionof{flisting}{Input\label{lst:input}}
\end{minipage}
\begin{minipage}{0.1\textwidth}
\ 
\end{minipage}
\begin{minipage}{0.4\textwidth}
\begin{lstlisting}
structure M1 : V {
	Num = {1..5}
	A = 4
	B = 5
}
structure M2 : V {
	Num = {1..5}
	A = 5
	B = 5
}
structure M3 : V {
	Num = {1..5}
	A = 5
	B = 4
}
\end{lstlisting}
\captionof{flisting}{Output\label{lst:output}}
\end{minipage}
\end{center}

\subsubsection{Progression($T$, $I_0$)}

The second inference, progression, is an inference that needs a linear time theory and a structure over the same vocabulary. 
Unlike model expansion, which creates a model with a full planning, progression generates the models of the next time step. 
We can choose the next state (a `snapshot') before continuing.
This way we can dynamically simulate a linear time theory step by step. 

A simulation has two parts. 
The first inference returns a list of possible initial states with \texttt{initialise(T, S)}. 
The second inference, \texttt{progress(T,S$_\texttt{i}$)}, returns a list of models.

The models of these two inferences have a special vocabulary: the single state vocabulary. This is a vocabulary where every time parameter in a predicate or function of the original vocabulary is projected away. The model represents the state of the simulation at the current time.

\section{\IDPD}
\begin{figure}
\includegraphics[width=\textwidth]{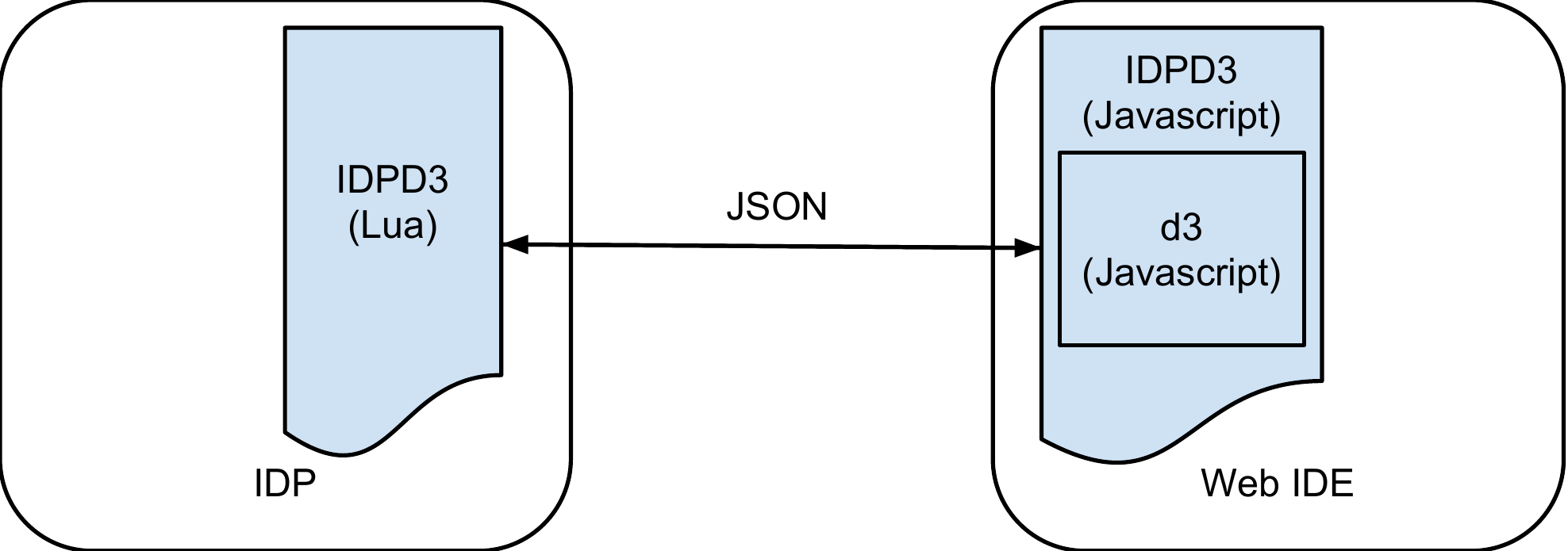}
\caption{The software environment that uses \IDPD \label{fig:idpd3stack}}
\end{figure}

\IDPD is a visualisation library for \IDP. It is the successor of \IDPDRAW~\cite{url:idpdraw}. 
The goal of this library is to easily visualise structures as a drawing. 
An example can be found in Figure~\ref{fig:lessenrooster}. 
\IDPD consists of two parts. The first part creates a description of a drawing from a structure as a JSON string, it is an encoding of the relevant predicates and functions. This transformation is done in the Lua environment of \IDP (cf. the left part of Figure~\ref{fig:idpd3stack}). 
The second part is integrated with the \IDPW~\cite{url:idp-IDE} and is written in Javascript. 
It interprets the description to visualise using the d3 visualisation library~\cite{2011-d3}. 
The d3 library can visualise any svg primitive that HTML supports. 
This part corresponds to the right part of Figure~\ref{fig:idpd3stack}.
Since Lua is always integrated in IDP and Javascript is a widely used language \IDPD is more platform independent than the previous version \IDPDRAW which is written in C++. 

To start, we will describe the features implemented in \IDPD. Next we will show how the encoding is done by the library and finally we will decode the input sent from the \IDPW.

\subsection{Features}

\IDPD can visualise structures in an interactive way, the current snapshot in a sequence of snapshots that correspond to a model of the linear time theory. 
This way it can visualise the output of many forms of inferences such as model expansion, minimisation, expanding linear time theories and the  simulation of linear time theories.
To support this \IDPDRAW introduces frames. 
These are multiple drawings that can be viewed like a slideshow. 

In \IDPD frames are extended with animated transitions that are built in d3. 
Elements are defined by their keys. 
When attributes change for an element with the same key between two different frames, they can be animated. 
This helps the viewer to track the changed elements, even when some frames are skipped.
For example the positions, sizes and colours can be interpolated by d3.
Other animations, like morphing a rectangle to a circle, are not implemented, but they could be easily added without modifying the JSON encoding. 

Elements can be one of four basic primitives: a rectangle, a circle, text or an image. 
Each has different attributes. 
Every primitive has a position (x, y), a z-order and a colour. 
Rectangles also have a size (width, height). 
Circles have a radius.
Text has a string (label) and a size. 
Images have a URL, a size(width, height) and no colour.

If the function value is missing for some key the value may default. Colours default to black. 
The other numeric attributes default to zero, this means that when no size is given, the element will be invisible. When no position is given, the element will be positioned at the edge of the visualisation.

There is also a special primitive to create an undirected visual link between two basic primitives. 
It has three attributes: link-from, link-to and link-width. 
The first two declare the keys to the primitives that must be linked. 
The third declares the width of the link that must be drawn.
This enables us to draw graphs just as Kara, \ASPVIZ and \IDPDRAW.
These tools also have automatic placement functions for graphs. 
In \IDPD this is done with a declaration whether a primitive is a node and whether the primitives position is fixed. 
Declaring as a primitive a node adds it to a force-directed layout implemented in d3.

Finally \IDPD adds a hook on the elements. When an element is clicked a JSON string is sent to the \IDP process. This string contains an identification of the current time-frame and the key of the element that was clicked. This hook is created to support simulated linear time theories. 

\subsection{Creating drawings}
For a visualisation we need a structure for which the vocabulary is an extension of the \IDPD output vocabulary.
The creation of this structure can be done in three ways. The first is to manually specify a structure. 
The second option is to place the visualisation code directly inside the original theory and expand this theory to a model.
The third option is to separate the two logic theories and expand the visualisation theory together with the generated model of the original theory as an extra step. 
The last approach allows us to separate the concern of the original model from the visualising model. This separation of concerns is one of the core values in software engineering.

In Listing~\ref{lst:idpd3_types} the types of the vocabularies, both for input and for output, are defined.
The output predicates and functions are also shown in this listing. 
These are the symbols that need to be used to create a visualisation.

\begin{center}
\begin{lstlisting}
vocabulary V_types {
  type shape constructed from {circ, rect, text, link, img}
  type time isa int
  type key isa string
  type color isa string
  type label isa string
  type width isa int
  type height isa int
  type order isa int
  type image isa string
}
vocabulary V_out {
  extern vocabulary V_types
    
  d3_width(time) : width
  d3_height(time) : height
  partial d3_type(time, key) : shape
  partial d3_x(time, key) : width
  partial d3_y(time, key) : height
  partial d3_color(time, key) : color
  partial d3_order(time, key) : order
  partial d3_circ_r(time, key) : width
  partial d3_rect_width(time, key) : width
  partial d3_rect_height(time, key) : height
  partial d3_text_label(time, key) : label
  partial d3_text_size(time, key) : width
  partial d3_img_path(time, key) : image
  partial d3_link_width(time, key) : width
  partial d3_link_from(time, key) : key
  partial d3_link_to(time, key) : key
  d3_node(time, key)
  d3_isFixed(time, key)
}
\end{lstlisting}
\captionof{flisting}{The types vocabulary \label{lst:idpd3_types} and the output vocabulary in \IDPD \label{lst:idpd3_out}}
\end{center}

\IDPD contains a procedure to transform a structure over the V\_out vocabulary to a JSON string.
The algorithm does this by looping over every \IDPD symbol in the output structure and reading every tuple of the predicate or of the graph interpretation of the function.
The function mapping is then added to the element with the correct time-frame and key as a key-value pair. 
Extending this algorithm with a new attribute is easy, it is a new symbol that must be read and the tuples must be transformed to a new key-value pair to add at the correct place.

When this JSON string is sent to the \IDPW it will be interpreted as an image and it will be visualised.
First it is parsed and then interpreted by the online part of the library. 
The JSON encoding was chosen in such a way that it coincides largely with the data-driven approach of the d3 library.

\subsubsection{Application: visualising a structure}
A structure over an extension of the output vocabulary can be visualised just by transforming it to JSON. 
For example the structure in Listing~\ref{lst:exampleStruct} has a rectangle with basic attributes a position (2,3), a width(4), a height(5) and with the default colour black. 
The dots signify places where more elements and attributes can be defined.

\begin{center}
\begin{minipage}{\textwidth}
\begin{lstlisting}
structure S : idpd3::V_out {
	d3_type = {1, "key", rect(); ...}
	d3_width = {1, "key", 4; ...}
	d3_height = {1, "key", 5; ...}
	d3_x = {1, "key", 2; ...}
	d3_y = {1, "key", 3; ...}
	...
}
procedure main() {
	visualise(S);
}
\end{lstlisting}
\captionof{flisting}{Example structure\label{lst:exampleStruct}}
\end{minipage}
\end{center}

This structure is transformed to the following JSON specification by Lua. This can be done at any time for a two-valued structure over V\_out by calling ``idpd3:visualise(structure)''. This method will transform the structure and print the specification. 
\begin{verbatim}
{"animation": [{"time":1, "elements": [{"key":"key", "type":"rect",
  "y":3, "x":2, "rect_height":"5", "rect_width":"4"}, ...]
, ...}, ...]}
\end{verbatim}

The \IDPW will filter the specification and visualise it immediately. Thus the \IDPW will draw this rectangle (and the other elements).

\subsubsection{Application: Transforming a model to a drawing}
As a general principle in software development different responsibilities should be divided as much as possible.
When you have a logic structure and theory that is expanded to a model you might want to visualise it.
Without changing the original theory you can create a visualising theory to expand the original model to a model that can be visualised.

For example the structure S is a three by three chessboard grid. 
The structure is expanded under the theory T to a structure sol and then visualised.
\begin{lstlisting}[mathescape=true]
vocabulary V {
  type X isa int
  isBlack(X,X)
}
structure S : V {
  X = {1..3}
  isBlack = {(1,2); (2,1); (2,3); (3,2)}
}
vocabulary V_out {
  extern vocabulary V
  extern vocabulary idpd3::V_out
  toKey(X, X) : key
}
theory T : V_out {
 {
  d3_type(1, toKey(x, y)) = rect.
  d3_x(1, toKey(x, y)) = 4*x - 2.
  d3_y(1, toKey(x, y)) = 4*y - 2.
  d3_rect_width(1, toKey(x, y)) = 4.
  d3_rect_height(1, toKey(x, y)) = 4.
  d3_color(1, toKey(x, y)) = "black" $\leftarrow$ isBlack(x, y).
  d3_color(1, toKey(x, y)) = "white" $\leftarrow$ $\lnot$isBlack(x, y).
  d3_width(1) = 14.
  d3_height(1) = 14.
 }
}
procedure toKey(x, y) {
    return x.."-"..y;
}
structure S_out : V_out {
    time = {1}
    color = {"black"; "white"}
    width = {1..15}
    height= {1..15}
    X = {1..3}
    toKey = procedure toKey
}
procedure main() {
  local m = merge(S, S_out);
	local sol = onemodel(T, m);
	visualise(sol);
}
\end{lstlisting}

\subsubsection{Application: Comparing theories}
Another of our basic applications checks whether two theories are (approximately) equivalent on some partial structure. 
It is designed to highlight the errors of the user (or student) designed theory compared to a given correct theory. 
The application needs five arguments: a theory the user created ($T_{user}$), the correct theory($T_{corr}$), a visualisation theory($T_{vis}$) and two structures that contain as basic information for the user ($S$) and the basic information for the visualisation ($S_{out}$). 

The application will try to find a model of the first theory that is not satisfied by the second theory. 
If it finds such a model exists, it is shown as a drawing where the errors are highlighted. 
Otherwise it will show one of the matching models.
To reduce the amount of computing time, a finite number of models can be checked. 

\begin{figure}
\includegraphics[width=\textwidth]{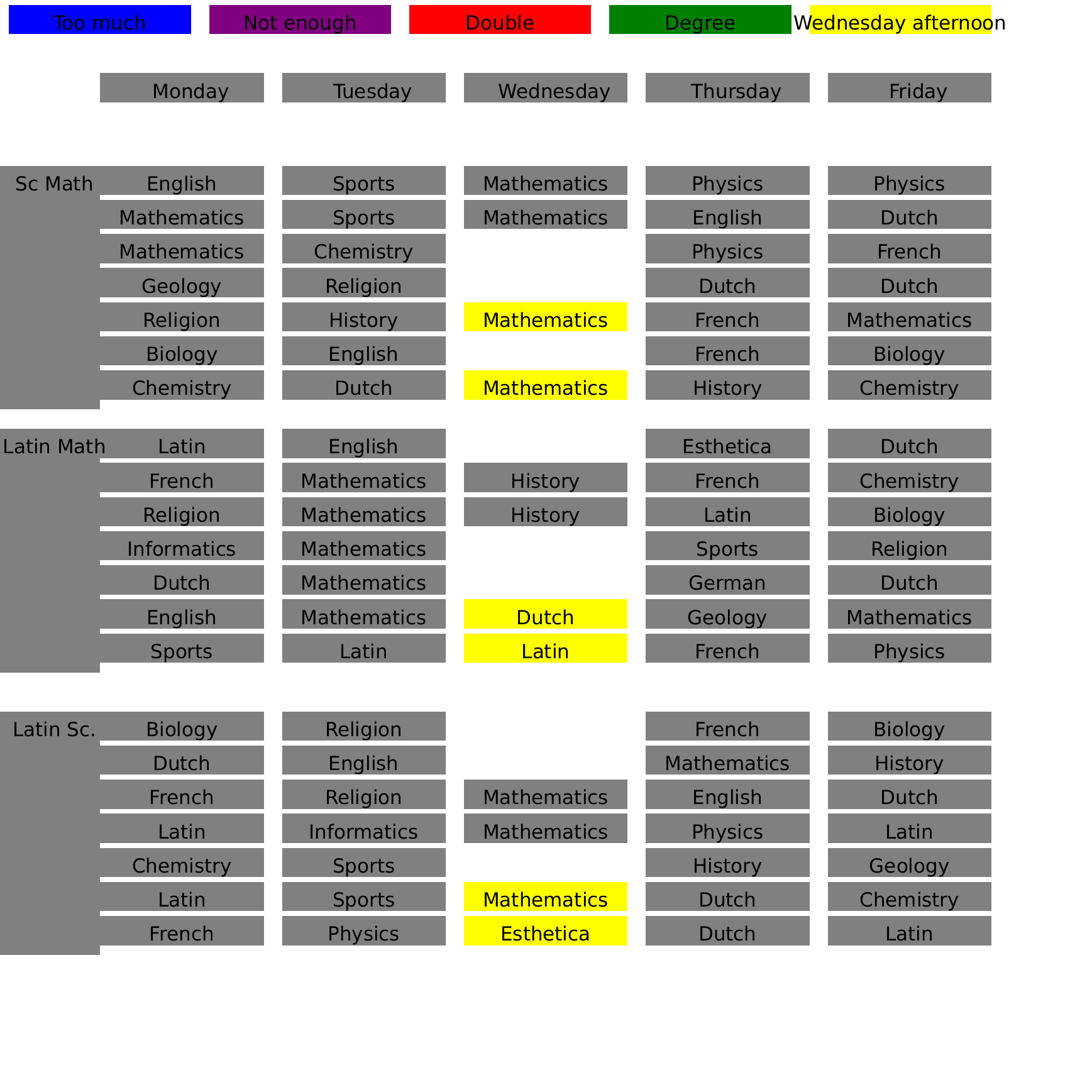}
\caption{An example of the images that can be produced with \IDPD \label{fig:lessenrooster}}
\end{figure}

This application is used in the examples of the online editor of \IDP. 
The example of Figure~\ref{fig:lessenrooster} can be found at \linkLesrooster. There are currently two rules that are wrongly encoded, which is displayed in the images as the rectangles which are coloured according to the error that is made. 
The correct rules are supplied in the comments, activating these will remove the coloured rectangles.

\subsection{Interpreting interactions}

Another extension of \IDPDRAW  in \IDPD is to support interaction with the user using logic LTC theories and progression inference. 
After an initial model is created and visualised, \IDP will wait for input from the \IDPW. 
This input is interpreted by the library and transformed to an input structure. 
The union of the current snapshot and the input structure is expanded to a structure that holds the current state and the chosen action. 
This structure is then used to progress to the next state.
This way we can create an interactive simulation of an LTC theory.
This loop behaves like the Model-View-Controller pattern. 
Where the output theory creates the View, the input theory behaves as the Controller, and the progression theory handles the Model.

In \IDPD the input is only click-based: the only predicate that is available in the input vocabulary is {\tt d3\_click(Time, Key)}.
The input is transformed to an \IDP structure by handling every object defined by a time-key pair. 
For example when an element with key ``key'' is clicked the drawing program will generate the following specification:
\begin{verbatim}
[{"time":1,"elements":[{"key":"key","type":"click"}]
\end{verbatim}

This is transformed to the following structure:
\begin{lstlisting}
structure S : V {
  time = {1}
  key = {"key"; ...}
  d3_click = {1, "key"}
}
\end{lstlisting}

\begin{figure}
\begin{minipage}{0.49\textwidth}
\begin{center}
\includegraphics[width=\textwidth]{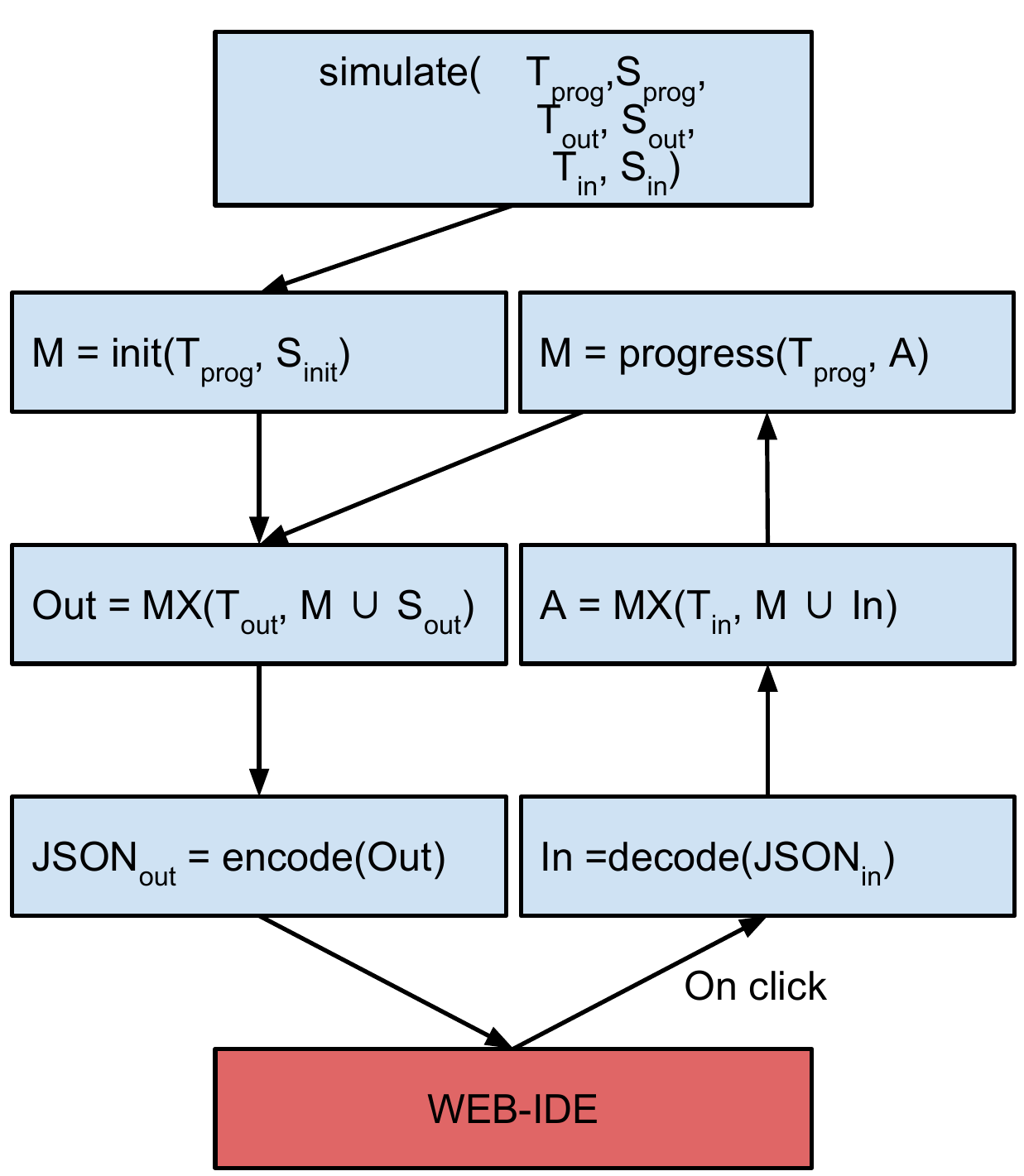}
\caption{The MVC loop\label{fig:MVC}}
\end{center}
\end{minipage}
\begin{minipage}{0.49\textwidth}
\begin{center}
\includegraphics[width=\textwidth]{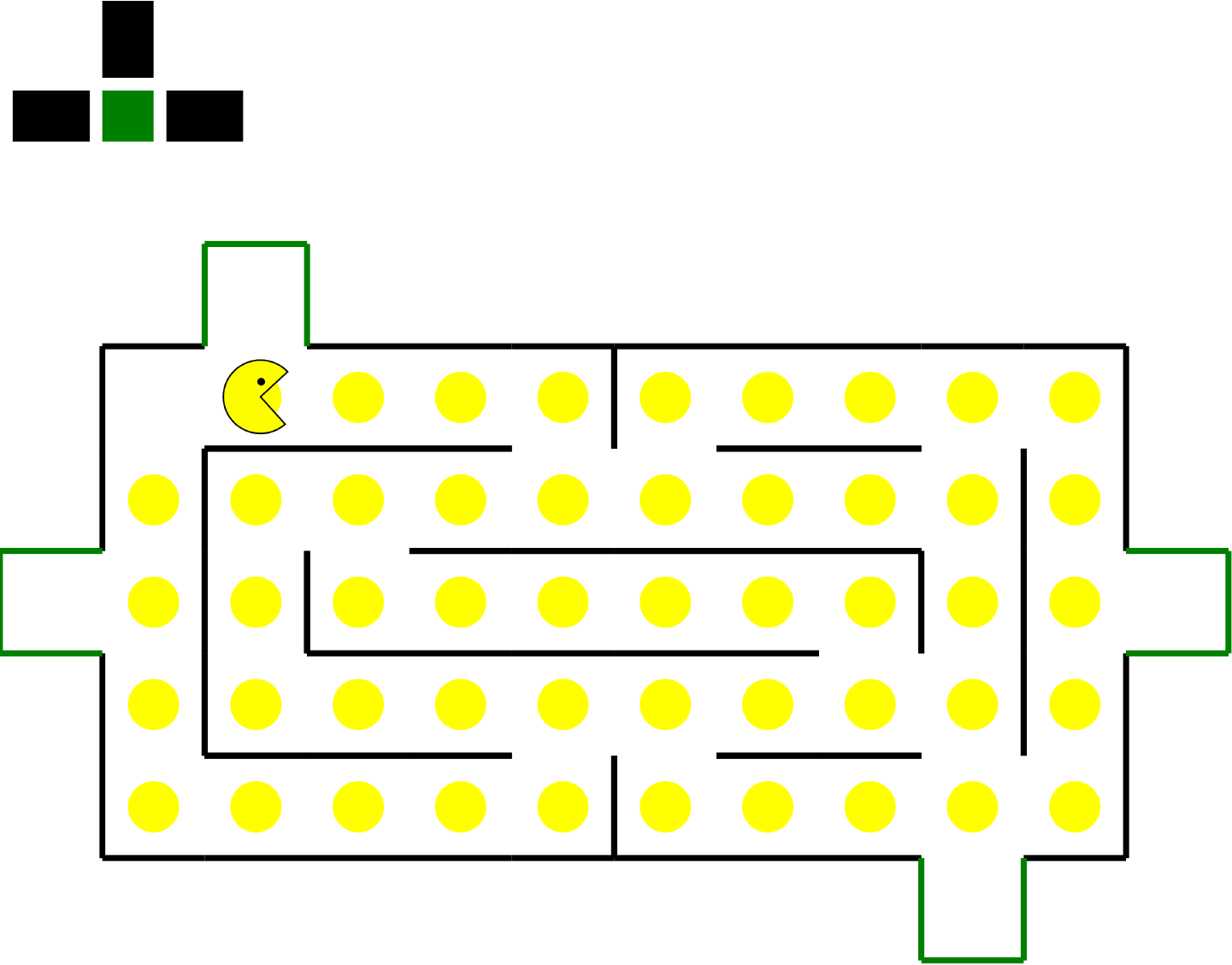}
\caption{Pac-Man as a game in \IDP\label{fig:pacman}}
\end{center}
\end{minipage}
\end{figure}

\subsubsection{Application: Interactive simulation}
Another application enables the interactive simulation of an LTC theory. This application needs 9 arguments. 
The first three are the progression theory $(T_{prog}$), the initial structure ($\struct_{init}$) and the output vocabulary ($V_{state}$). 
The next three handle the output: the output theory ($T_{out}$), structure ($S_{out}$) and vocabulary ($V_{out}$). 
The last three handle the input: the theory ($T_{in}$), structure ($\struct_{in}$) and vocabulary ($V_{in}$).

This application first calculates an initial model $M_0$ by expanding 
$S_{init}$ with the theory $T_{prog}$. 
This model is merged with the output structure $S_{out}$ and expanded to the visualisation model $M_{v0}$ using $T_{out}$. 
This visualisation model is transformed using the library to a JSON string and visualised by the \IDPW.
The application then waits for input.

When input is received it is added to $S_{in}$ creating a more specific structure. 
This structure is expanded with $T_{in}$ and projected to the actions of the model. 
The application keeps looping until no new model exists for one of the inferences. 
A schematic view of the interactions between the theories and the structures is given in Figure~\ref{fig:MVC}.

In this way an interactive game, like Pac-Man (Figure~\ref{fig:pacman}), can be made. 
This application is available at \linkPacMan.

\section{Full example}
Suppose we have an LTC theory with a single function, count(Time) : Count. There are also three actions: countUp(Time), countDown(Time), setValue(Time, Value). This theory would be declared the following way:

\begin{lstlisting}[mathescape=true]
LTCvocabulary V_types {
	type Count isa int
	type Time
	Next(Time) : Time
	Start : Time
}
LTCvocabulary V_state { ... }
LTCvocabulary V_action { ... }
LTCvocabulary V { ... }

theory T : V {
 {
  count(Start) = 0.
  count(Next(t)) = v $\leftarrow$ setValue(t, v).
  count(Next(t)) = count(t) + 1 	$\leftarrow$ countUp(t).
  count(Next(t)) = count(t) - 1 	$\leftarrow$ countDown(t).
  count(Next(t)) = count(t) $\leftarrow$ $\lnot$countUp(t) $\land$
		$\lnot$countDown(t) $\land$ $\forall$v : $\lnot$setValue(t, v).
 }
}
\end{lstlisting}

This theory can be augmented with \IDPD by adding the visualisation theories. 
The output vocabulary is the union of the single-state vocabulary\footnote{V\_ss is an automatically generated vocabulary where Time is projected away from the vocabulary.} and the \IDPD output-vocabulary. 
Helper functions and predicates can be added in this vocabulary.
The output theory declares two text elements: the current count and a label to count up.
The input vocabulary is the union of the single-state vocabulary and the \IDPD input-vocabulary. 
The input theory declares two rules to convert the clicks from the user: when the ``Count up'' text is clicked the counter is incremented, when the count itself is clicked the counter resets to zero.

\begin{lstlisting}[mathescape=true]
LTCvocabulary V_d3 {
    extern vocabulary V_types
    extern vocabulary idpd3::V_types    
    toLabel(Count) : label
    countLabel : label
}
vocabulary V_d3_out { ... }
vocabulary V_d3_in { ... }

theory T_out : V_d3_out {
 {
  d3_width(1) = 10.
  d3_height(1) = 10.

  d3_type(1, "label") = text.
  d3_x(1, "label") = 1.
  d3_y(1, "label") = 1.
  d3_text_size(1, "label") = 1.
  d3_text_label(1, "label") = toLabel(count).
  
  d3_type(1, "button") = text.
  d3_x(1, "button") = 1.
  d3_y(1, "button") = 5.
  d3_text_size(1, "button") = 1.
  d3_text_label(1, "button") = "Count up".
}}
theory T_in : V_d3_in {
 {
  countUp $\leftarrow$ d3_click(1, "button").
  setValue(0) $\leftarrow$ d3_click(1, "label").
}}
\end{lstlisting}

Finally the basic structure and the Lua-scripting environment are declared.  An \IDP procedure is used to automatically convert from a number to a string. 
And due to quirks in \IDP with types there is one structure that is projected to the different structures needed for the application. 
It is possible to disambiguate it, at the cost of using more advanced features in \IDP. The full source code is available for testing at \linkMinimalExample.
\begin{lstlisting}
structure S : V {
    Count = {0..100}
    Start = 0
}
structure S_d3 : V_d3_ss {
    Count = {0..100}
    time = {1}
    key = {"label"; "button"}
    //label is autogenerated by procedure output
    width = {0..20}
    height = {0..20}
    countLabel = "Count up"
    toLabel = procedure toText
}
procedure main() {
    stdoptions.splitdefs = false;
    stdoptions.postprocessdefs = false;
    stdoptions.cpsupport = false;
    stdoptions.xsb = false;
    idpd3_browser:setLogLevel(0);
    local go = idpd3_browser:createLTC(
        T, S, V_state_ss,
        T_in, S_d3, V_d3_in,
        T_out, S_d3, V_d3_out);
    local lastState = go();
}
\end{lstlisting}

\section{Related work}
Multiple systems have already been created to visualise logic programs. The main systems are \ASPVIZ~\cite{cliffe2008aspviz}, \IDPDRAW~\cite{url:idpdraw} and Kara~\cite{kloimullnerkara}.

\ASPVIZ is one of the first visualisation tools for logic programs. 
This program joins the original logic program with a logic visualisation program before expanding it with an ASP solver. 
As it is one of the early visualisation tools created, it only supports basic primitives and visualising multiple frames. \ASPVIZ supports saving images as svg, like \IDPD and Kara.

\IDPDRAW is another visualisation tool. It is the predecessor of \IDPD.
The main improvement over \IDPDRAW is that the program is built on the core of \IDP itself. 
We use an \IDP vocabulary, which helps during the construction of visualisation theories with grammar checking and debugging. 
Additionally, the transformation of the structures are written in Lua, the imperative language that drives the inferences. 
This means that while \IDPDRAW is capable of providing interactivity, it is limited to direct keyboard interaction with \IDP and spawning multiple windows. \IDPD has extended this to clicking with the mouse on drawn items.

One of the newest visualisation tool is Kara.
This system is part of the SeaLion IDE. 
It supports an almost full superset of both \ASPVIZ and \IDPDRAW, lacking only the ability to show short non-interactive frames. 
Like \IDPDRAW and \IDPD, Kara supports ordering elements by a z-axis.
One of the strengths of Kara is that it supports some higher-level specifications that are not supported by the other main systems. 
It can generate layouts for graphs and visualise an arbitrary answer set model as a hyper-graph. 

\section{Future work}
In the future we would like to extend the framework of interactive animations to more practical applications. 
For this some basic functionality must be added: keyboard interaction and text fields are a standard way of entering information.

Another change we would like to implement is generalising the vocabulary of both input and output. 
Currently it takes a minimal but non-zero time to implement a new attribute.
If this is done new attributes can be added without changing the library.
This would decrease the time invested in maintaining the library in the same way as argued for the d3 visualisation library~\cite{2011-d3}.

However to really support practical applications we should need to move away from visualisations and support forms.
A new library that uses many of the ideas currently implemented in \IDPD, would be created for this.

\section{Conclusion}
In this paper we presented \IDPD as the successor of \IDPDRAW. 
The main feature of \IDPD  is the possibility to visualise an interactive simulation of a linear time theory.
Smaller features include the animation of basic primitives and the tutorial application.
Due to it's integration with both \IDP as a library and the \IDPW developing visualisations for \IDP is easier and platform independent.
The maintenance of the library should be easier than before.
In the future we would like to implement the library for more practical applications involving data entry and perhaps forms.

\printbibliography
\end{document}